\documentclass{PoS}

\newcommand{\beq}{\begin{equation}}
\newcommand{\eeq}{\end{equation}}
\newcommand{\bea}{\begin{eqnarray}}
\newcommand{\eea}{\end{eqnarray}}
\newcommand{\ba}{\begin{array}}
\newcommand{\ea}{\end{array}}
\newcommand{\bi}{\begin{itemize}}
\newcommand{\ei}{\end{itemize}}
\newcommand{\bn}{\begin{enumerate}}
\newcommand{\en}{\end{enumerate}}
\newcommand{\bc}{\begin{center}}
\newcommand{\ec}{\end{center}}

\newcommand{\ol}{\overline}

\newcommand{\De}{\Delta}
\newcommand{\Om}{\Omega}
\newcommand{\al}{\alpha}

\newcommand{\la}{\lambda}

\newcommand{\si}{\sigma}

\newcommand{\GeV}{\mathinner{\mathrm{GeV}}}

\title{Hidden sector dark matter and Higgs physics}

\ShortTitle{Hidden sector dark matter and Higgs physics}

\author{Seungwon Baek, \speaker{Pyungwon Ko}, 
Wand-Il Park and Eibun Senaha\\
        School of Physics, KIAS, Seoul 130-722, Korea\\
        E-mail: \email{pko@kias.re.kr}}

\abstract{
We consider a hidden sector dark matter, where a singlet fermion is 
a cold dark matter and a real singlet scalar boson $S$ is a messenger 
between the SM and the hidden sectors. 
This singlet scalar will mix with the SM Higgs boson $h$, 
and we expect there are two Higgs-like scalar bosons $H_1$ and $H_2$.
Imposing all the relevant constraints from collider search bounds on Higgs 
boson, DM scattering cross section on proton and thermal relic density. 
We find that there is a destructive interference between $H_1$ and $H_2$ contributions
to the direct detection cross section of the DM.  Also one of the two Higgs-like scalar 
bosons can easily escape the  detections at the LHC, and there will be a universal
reduction of the signal strength for the observed 125 GeV Higgs-like boson, which 
could be tested at the LHC with more data in the future. 
}

\FullConference{VIII International Workshop on the Dark Side of the Universe,\\
		June 10-15, 2012\\
		Rio de Janeiro, Brazil}

\begin{document}


\section{Motivations}

The standard model (SM) has been tested by many experiments from 
atomic scale up to electroweak scale, and was extremely successful. 
However, there are three observational facts which call for new physics beyond the SM:
\begin{itemize}
\item Baryon number asymmetry of the universe (BAU)
\item Neutrino masses and mixings
\item Nonbaryonic Cold dark matter (CDM).
\end{itemize}
There are many models and suggestions for each problem listed above.
The most economical way to solve the BAU and the neutrino masses and 
mixings is probably to introduce heavy right-handed neutrinos and invoke 
seesaw mechanism. Leptogenesis can turn to baryogenesis around electroweak 
phase transition by sphalerons. 

Given the triumphant success of the SM with its aforementioned shortcomings 
kept in mind,  the most important questions in particle physics at the LHC era 
would be the following: 
\begin{itemize}
\item Origin of electroweak symmetry breaking (EWSB)
\item Nature of cold dark matter (CDM) 
\item Origin of flavor structure and families.
\end{itemize}

Any new physics models at the electroweak scale are strongly constrained
by electroweak precision test and CKM phenomenology, if new physics 
feels the SM gauge interactions. In this case, the new physics scales 
should be larger than $O(1)$ TeV and $O(100)$ TeV in order to be safe 
from the EWPT and CKM phenomenology, respectively. 
On the other hand, fine tuning problem of $($Higgs mass$)^2$ requires
that new physics scale should be around $\lesssim O(1)$ TeV. 
Thus there is strong tension between two.

On the other hand, if new physics particles are neutral under the SM 
gauge group and do not feel SM gauge interactions, the constraints from
the EWPT and CKM fit can be relaxed by a significant amount, and 
new physics scales could be easily at the EW scale. Thus we are led to consider 
a weak scale hidden sector which is neutral under the SM gauge interaction.
The hidden sector matters can be natural CDM if there are suitable messengers
between the SM and the hidden sectors. 
In this talk, I will discuss singlet fermion hdden sector dark matters based on the
works  \cite{ko2012,ko2012_2}. (Another interesting case where the hidden sector 
gauge interaction is strong and confining like the ordinary QCD is discussed in 
Ref.s~ \cite{ko2007,ko2008,ko2011}.)
In this approach, there is no resolution of fine tuning problem of $($Higgs 
mass$)^2$ parameter, since new particles do not carry the SM gauge 
quantum numbers. 
And I don't address the fine tuning problem in this talk. 

\section{Hidden sector singlet fermion DM model}
\subsection{Model}
Let us consider a singlet fermion dark matter $\psi$ with a real singlet 
scalar messenger $S$ with the following lagrnagian \cite{ko2012}:
\bea
 {\cal L} = {\cal L}_{\rm SM} + {\cal L}_{\rm hidden} + {\cal L}_{\rm portal},
\label{eq:Lag}
\eea
where 
\bea
{\cal L}_{\rm hidden} &=& {1 \over 2} (\partial_\mu S \partial^\mu S - m_S^2 S^2) 
-\mu_S^3 S - {\mu_S^\prime \over 3} S^3  - {\la_S \over 4} S^4 + 
\ol{\psi} ( i \not \partial - m_{\psi_0} ) \psi 
- \la S \ol{\psi} \psi,  
\\
{\cal L}_{\rm portal} &=& - \mu_{HS} S H^\dag H -{\la_{HS} \over 2} S^2 H^\dag H
,
\label{eq:Lag2}
\eea
We assume $\psi$ carries a conserved dark charge, and is distinguished from
the right-handed neutrinos. 

In the literature, the Higgs portal fermion dark matter is often discussed using the 
following effective lagrangian : 
\beq
{\cal L}_{\rm portal}  = 
- \overline{\psi} \left( m + \lambda_{\psi H} \frac{H^\dagger H}{\Lambda} \right) \psi
\eeq
There are two scalar bosons in our model ($h$ and $s$), and we will find that 
the physics results from our model are very different from those based on Eq.~(2.4). 

\subsection{Constraints}

We consider the following constraints on the model parameters: 
\begin{itemize}
\item the perturbative unitarity condition on the Higgs 
sector 
\item the LEP bound on the SM Higgs boson mass 
\item the oblique parameters $S$, $T$ and $U$ obtained from the EWPT
\item the observed CDM density, 
$\Omega_{\rm CDM} h^2 =0.1123 \pm 0.0035$  
which we assume is saturated by the thermal relic $\psi$,
\item the upper bound on the DM-proton scattering cross section 
obtained by the XENON100 experiment.  
\end{itemize}
Note that the first three constraints are independent of the dark matter sector,
and they apply to the SM plus a singlet scalar model without dark matter as well.

The extended Higgs sector gives extra contribution to the gauge boson 
self-energy diagrams, as the SM Higgs boson does.
This can affect the EWPT leading to  the constraints on the oblique parameters, 
$S, T$ and $U$, by the Higgs sector. 
It turns out that the EWPT constraint on our model is 
generically much less severe than on the SM. 
The SM always predicts a negative $\De T$ for the Higgs mass larger than 
$m_h = 120$ GeV. 
However, $\De T$ can be either positive or negative in our model, and the mixing
between the singlet scalar is fine with the EWPT (see Fig.~1). 

\begin{figure}
\centering
\includegraphics[width=0.5\textwidth]{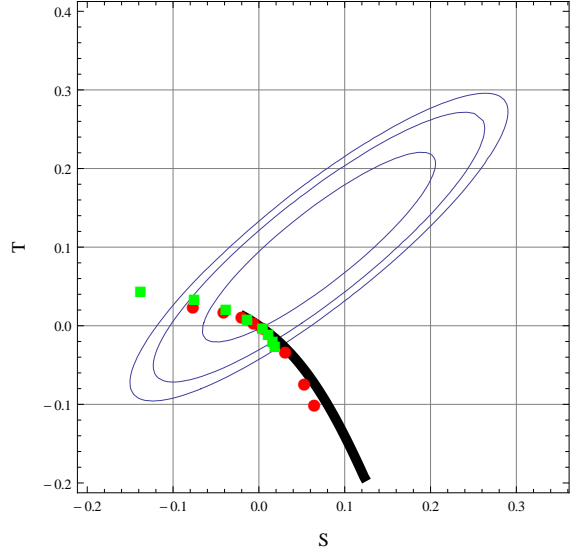}
\caption{
The prediction of $(S,T)$ parameters. We fixed the reference Higgs mass
to be 120 GeV.
The ellipses are (68, 90, 95)~\% CL contours from the global fit.
The thick black curve shows the SM prediction with the Higgs boson mass 
in the region $(100,720)$ GeV.
The red, green dots correspond to $\alpha=45^\circ, 20^\circ$, respectively.
The dots are for the choices $(m_1,m_2)(\GeV) =(25,125),(50,125),(75,125),(100,125),
(125,125),(125,250),(125,500),(125,750)$ from above for each color.
}
\label{fig:ST125}
\end{figure}

\subsection{Dark matter phenomenology}

\begin{figure}
\centering
\includegraphics[width=0.6\textwidth]{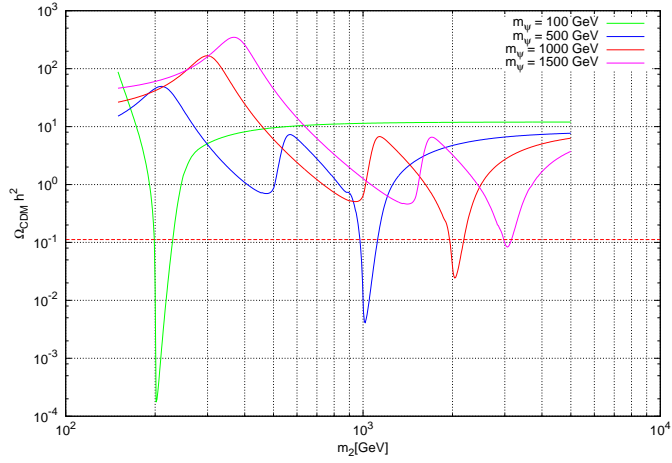}
\caption{Dark matter thermal relic density ($\Omega_{\rm CDM} h^2$) 
as a function of $m_2$ for $m_1=125 \GeV$, $\lambda=0.4$, $\alpha=0.1$ 
and $m_\psi = 100, 500, 1000, 1500 \GeV$ from top to bottom at right side.
The dotted red line corresponds to the observed value, $\Omega_{\rm CDM} h^2 = 0.112$.}
\label{fig:relic-density}
\end{figure}

The observed DM relic density,
$\Omega_{\rm CDM} h^2 \simeq 0.1123 \pm 0.0035$ 
is related with the thermally averaged annihilation cross section times relative
velocity at freeze-out temperature. 
The annihilation cross section of a DM pair is proportional to 
$\sin^2 2 \alpha$. Since the EWPT and LHC observation of the SM-like Higgs boson
restricts $\alpha$ to be small, the cross section is generically much smaller
than is needed to explained the current relic density. This can be seen in
Fig.~\ref{fig:relic-density} except for resonance regions.

We used the micrOMEGAs package 
for numerical calculation of DM relic density and direct detection cross section. 
In Fig. \ref{fig:relic-density}, we show the CDM relic 
density as  a function of $m_2$ for various choices of $m_\psi = 100, 500, 
1000, 1500$ GeV, with $\lambda = 0.4$ and $\alpha = 0.1$. 
We can always find out the $m_2$ value which can accommodate thermal  
relic density of the singlet fermion CDM $\psi$. Note that there is no strong
constraint on the heavier Higgs with a small mixing angle $\alpha$, because 
$H_2$ would be mostly a singlet scalar so that it is very difficult to produce 
it at colliders, and also it could decay into a pair of CDM's with a substantial
branching ratio.

The dark matter scattering on proton target is given by 
\bea \label{d-sigma-th}
\sigma_p 
&\simeq& 8.6 \times 10^{-9} \ {\rm pb} \left( \frac{125 \GeV}{m_1} \right)^4 
\left( 1 - \frac{m_1^2}{m_2^2} \right)^2 \left( \frac{\lambda \sin \alpha 
\cos \alpha}{0.1} \right)^2 .
\eea
Note that there is a generic cancellation between the $H_1$ and $H_2$ 
contributions.  Due to this cancellation,  the constraint from direct detection of 
dark matter becomes much weaker on the Higgs couplings to
the DM's.  If we ignored the singlet scalar from the beginning (for example in the 
effective lagrangian approach using Eq. (2.4)), 
we could not enjoy such cancellation in $\sigma_p$, so that the Higgs 
coupling to the DM would be much more tightly constrained.

\subsection{Vacuum stability}
Compared with the SM, there are additional fields $\psi$ and $S$ that contribute to the
effective potential for $H$ and $S$, and the vacuum stability can be modified from the SM 
case. We studied this issue very carefully in Ref.~\cite{ko2012_2}, and found that the
SM vacuum can be stable up to Planck scale within this model (see Fig.~3).

\begin{figure}[htbp]
\begin{center}
\includegraphics[width=0.45\textwidth]{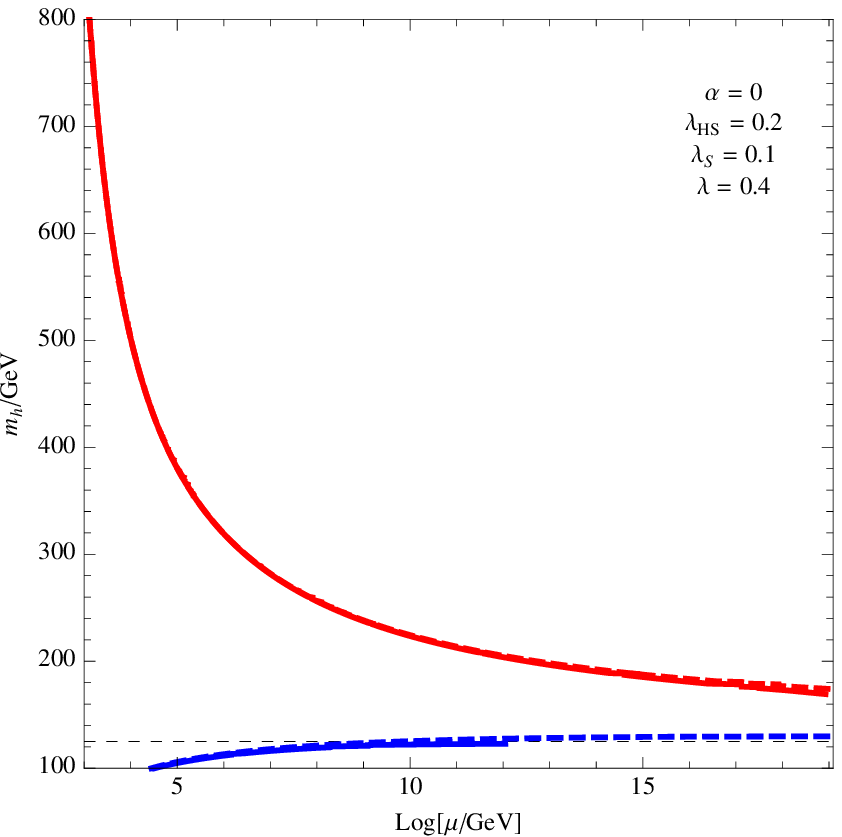}
\includegraphics[width=0.45\textwidth]{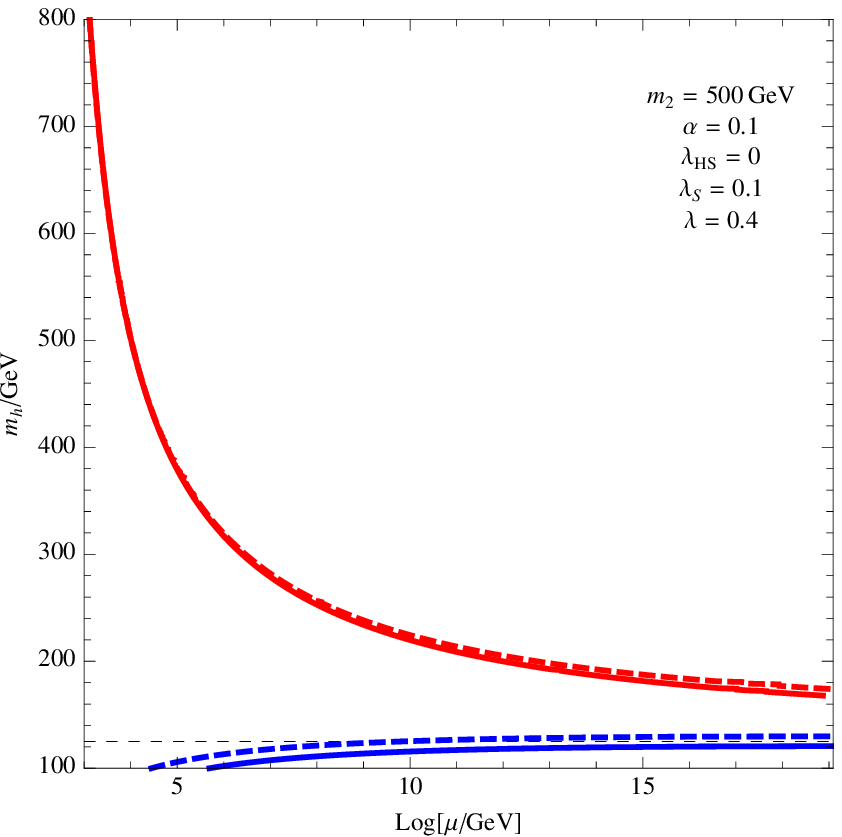}
\caption{The mass bound of SM-like Higgs ($m_1$) as a function of energy scale for $(\alpha,\lambda_{HS})=(0,0.2)$(left),$(0.1,0)$(right) with $\lambda_S = 0.1$ and $\lambda = 0.4$.
The red/blue line corresponds to triviality/vacuum-stability bound in SM(dashed) and our model(solid).
The dashed black line corresponds to $m_1 = 125 \GeV$.
}
\label{fig:mh-vs-mu}
\end{center}
\end{figure}

\section{Implications for the Higgs search at the LHC}

In this section, we investigate  if it is possible to discover the  Higgs(es) at the LHC,
taking into account of all the constraints discussed in the previous section.   
The signal strength or  ``the reduction factor'' in the event number 
of a specific final state SM particles,  $X_{\rm SM}$, in the Higgs 
boson decays is defined as
\beq
r_i \equiv \frac{\si_{H_i} B_{H_i \to X_{\rm SM}}}{\si^{\rm SM}_{H_i} 
B^{\rm SM}_{H_i \to X_{\rm SM}}}\ \ (i=1,2),
\eeq
where $\si_{H_i}$ and $B_{H_i \to X_{\rm SM}}$ are the production cross 
section of $H_i$, and the branching ratio of $H_i \to X_{\rm SM}$ respectively, 
while $\si^{\rm SM}_{H_i}$ and $B^{\rm SM}_{H_i \to X_{\rm SM}}$ are 
the corresponding quantities of the SM Higgs with mass $m_i$.
Note that the signal strength $r_i$ becomes less than "1"  
due to the mixing between $h$ and $s$, even if the invisible mode 
($H_i \rightarrow \psi \overline{\psi}$) or the Higgs-splitting  mode 
($H_2 \to H_1 H_1$) is kinematically forbidden in the Higgs decay.
In other words, a reduced signal of the Higgs boson at the LHC would 
be a generic signature of the mixing of the SM Higgs boson with 
extra singlet scalar boson(s).

We study the following three benchmark scenarios classified 
according to the Higgs mass relations:
\begin{itemize}
\item Scenario 1 (S1): $m_1 \sim 125$~GeV $\ll m_2$ 
\item Scenario 2 (S2): $m_1 \sim m_2 \sim 125$~GeV
\item Scenario 3 (S3): $m_1 \ll m_2 \sim 125$~GeV
\end{itemize}
We scanned the remaining parameters in the range
\begin{equation}
  0 < \la < 1, ~~~
 10 \ \ {\rm GeV} < M_\psi <100 \ \ {\rm GeV}, ~~~
 0 < \al < \pi/2 .
\end{equation}
All the points in the plots satisfy the constraints described earlier.

We can divide the $\si_p$ (in pb) into two regions:
\bea
 \si_p^>: 10^{-9} <  \sigma_p < 10^{-8}, \quad
 \si_p^<: \sigma_p < 10^{-9}, 
\eea
where the former region can be probed in near future direct search
experiments. The relic density is also divided into two regions:
\bea
 (\Om_{\rm CDM} h^2)^{3 \si}: 0.1018 < \Om_{\rm CDM} h^2 < 0.1228, \quad
 (\Om_{\rm CDM} h^2)^{<}:  \Om_{\rm CDM} h^2 < 0.1018.
\eea
where the former is the WMAP 3$\sigma$ allowed region.

The region that the LHC at 14 TeV can probe at 3$\sigma$ level  with 
5 (10) fb$^{-1}$ luminosity is represented by solid (dashed) line.
The S1 scenario can be tested fully at the LHC with 10 fb$^{-1}$
by observing $H_1$. 
In the case of S2 the LHC may see both Higgs bosons with the standard 
search strategy.
However, there are still some points which the LHC has difficulty to 
find two Higgs bosons.  These are the points near the origin 
($r_1 \approx r_2 \approx 0$)  where the invisible decays becomes dominant.
In S3 the region with small $r_2 (<0.24)$ can not be probed with the standard
decay channels. However, once $H_2 \to H_1 H_1$ is open, 
this region can also be tested at the LHC. 

\begin{figure}
\centering
\includegraphics[width=0.6\textwidth]{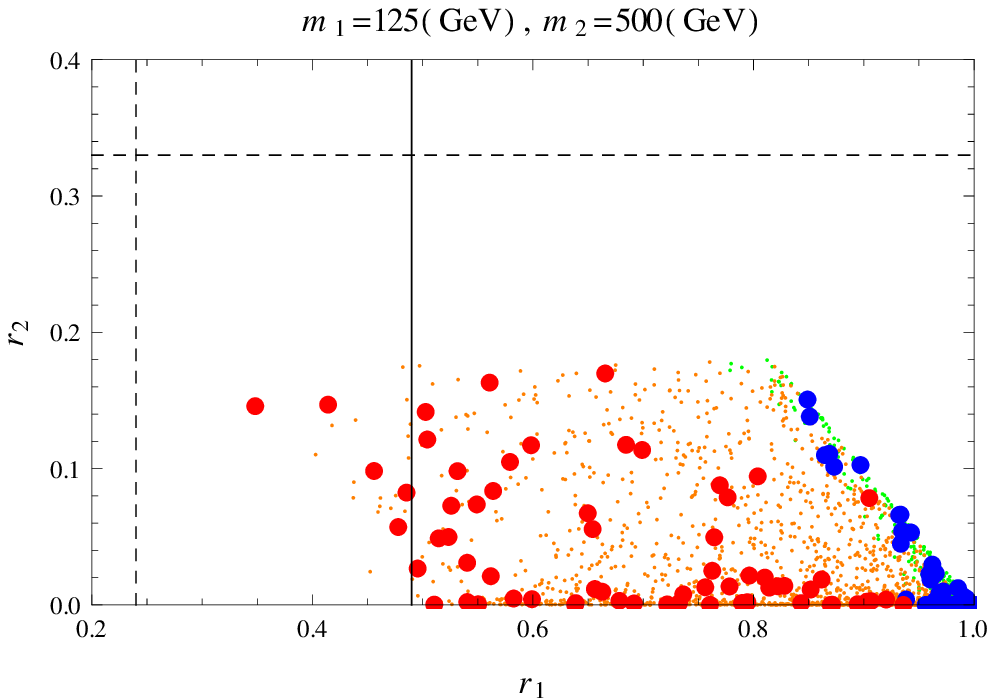}
\includegraphics[width=0.6\textwidth]{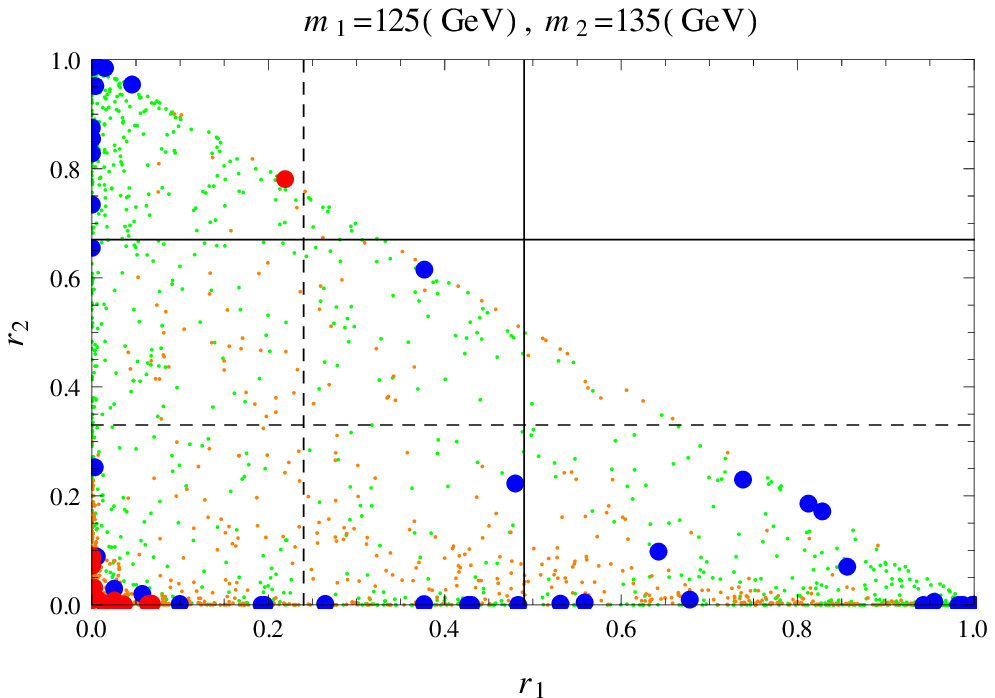}
\includegraphics[width=0.6\textwidth]{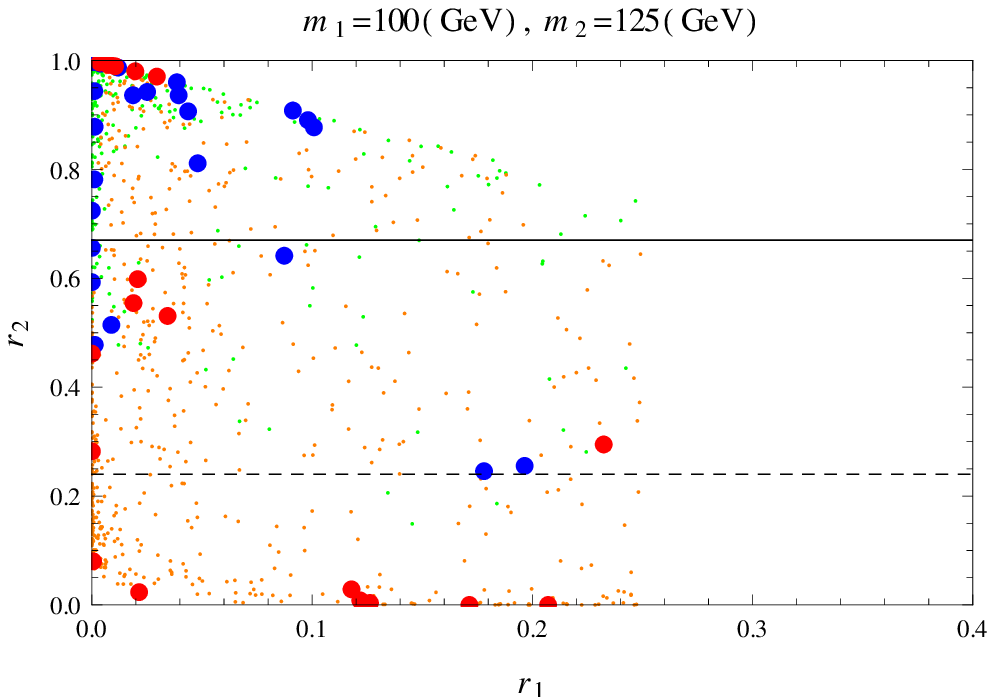}
\caption{Scatter plot in $(r_1,r_2)$ plane for the scenario 
S1, S2 and S3 (from above). The region that the LHC can probe 
at 3$\sigma$ level  with 5 (10) fb$^{-1}$ 
luminosity is represented by solid (dashed) line. 
The points represent 4 different cases:
$(\Om_{\rm CDM} h^2)^{3 \si}$, $\si_p^>$ (big red),
$(\Om_{\rm CDM} h^2)^{3 \si}$, $\si_p^<$ (big blue),
$(\Om_{\rm CDM} h^2)^{<}$, $\si_p^>$ (small orange),
and $(\Om_{\rm CDM} h^2)^{<}$, $\si_p^<$ (small green). 
}
\label{fig:r1r2}
\end{figure}

\section{Conclusions}

In this talk, I discussed an example of hidden sector dark matter model with a singlet 
fermion as a CDM, and discussed their interplay with phenomenology of the SM Higgs boson. 
Our results are completely different from those based on the effective lagrangian (2.4). 
The same is true for the Higgs portal vector dark matter~\cite{ko2012_3}. 

Generic signatures of hidden sector fermion dark matter can be 
summarized as follows:
\begin{itemize}
\item Thermal relic density of hidden sector DM can be easily compatible with the 
WMAP observation, and they can be detected in the direct detection experiments.
\item A real singlet scalar boson $S$ should be introduced as a messenger between the 
hidden sector and the SM sector, if the hidden sector has fermions and gauge bosons only. Therefore there are two physical Higgs-like scalar bosons. 
\item There is a destructive interference in the contributions from two scalar bosons
in direct detection cross section, which can not be seen in the effective lagrangian 
approach based on Eq.~(2.4). 
\item Higgs can decay into a pair of CDM, if kinematically allowed, which is begun to constrained
by the LHC data.
\item Production cross section for Higgs boson is smaller than the SM 
Higgs boson because of the mixing with composite scalars from the 
hidden sector. 
\item Depending on the parameters, only one or none of the two Higgslike
scalar boson(s) could be found at the LHC.
\item Recent results on the Higgs-like new boson 
with mass around with 125 GeV from the LHC will 
constrain this class of models. In particular there is a universal reduction of the signal
strength in all the channels. If the future data do not respect this universal suppression, 
our model would be excluded, independent of discovery of the second Higgs boson.
\end{itemize}


\section{Acknowledgements}
The author is grateful to Seungwon Baek, Taeil Hur, Dong Won Jung,  Jae Yong Lee 
and Wan-Il Park  and Eibun Senaha for enjoyable collaborations on the subjects 
reported in this talk and other related issues.  


\begin{thebibliography}{99}
\bibitem{ko2012}
S.~Baek, P.~Ko and W.~-I.~Park,
  JHEP {\bf 1202}, 047 (2012)
  [arXiv:1112.1847 [hep-ph]].

\bibitem{ko2012_2}
S.~Baek, P.~Ko, W.~~I.~Park  and E.~Senaha,
  JHEP {\bf 1211}, 116 (2012)
  [arXiv:1209.1685 [hep-ph]].
  
  \bibitem{ko2007}
T.~Hur, D.~-W.~Jung, P.~Ko and J.~Y.~Lee,
  Phys.\ Lett.\ B {\bf 696}, 262 (2011)
  [arXiv:0709.1218 [hep-ph]].

\bibitem{ko2008} 
  P.~Ko, Invited talk at the 4th International 
Conference on Flavor Physics, KITPC, Sep. 24-28, 2007, Beijing.
  Int.\ J.\ Mod.\ Phys.\ A {\bf 23}, 3348 (2008)
  [arXiv:0801.4284 [hep-ph]]; 
  P.~Ko, Invited talk at the Dark Side of the Universe, Melbourne.
  AIP Conf.\ Proc.\  {\bf 1178}, 37 (2009);
  Talk at ICHEP 2010, Paris. PoS ICHEP {\bf 2010}, 436 (2010)
  [arXiv:1012.0103 [hep-ph]].

\bibitem{ko2011} 
  T.~Hur and P.~Ko,
  Phys.\ Rev.\ Lett.\  {\bf 106}, 141802 (2011)
  [arXiv:1103.2571 [hep-ph]].



\bibitem{ko2012_3}
S.~Baek, P.~Ko, W.~~I.~Park  and E.~Senaha, 
"Higgs portal vector dark matter : revisted,"  in preparation.

\end{thebibliography}
\end{document}